\documentclass[lettersize,journal]{IEEEtran}

\usepackage{cite}
\usepackage{amsmath,amssymb,amsfonts}
\usepackage{algorithmic}
\usepackage{graphicx}
\usepackage{amssymb,comment}
\usepackage{algorithmic}
\usepackage{algorithm}
\usepackage{caption} 
\usepackage{subcaption}
\usepackage{scrextend}
\usepackage{verbatim}
\usepackage{color}
\usepackage{enumitem,epstopdf}

\usepackage{setspace}

\begin{document}

\title{Integrating Movable Antennas and Intelligent Reflecting Surfaces (MA-IRS): Fundamentals, Practical Solutions, and ISAC}
\author{Qingqing~Wu, Ziyuan~Zheng, Ying~Gao, Weidong~Mei, Xin~Wei, Wen~Chen, and Boyu~Ning \vspace{-12pt}
\thanks{Q. Wu, Z. Zheng, Y. Gao, and W. Chen are with the Department of Electronic Engineering, Shanghai Jiao Tong University, 200240, China (e-mail: \{qingqingwu, zhengziyuan2024, yinggao, wenchen\}@sjtu.edu.cn).}
\thanks{W. Mei, X. Wei, and B. Ning are with the National Key Lab of Wireless Communications, University of Electronic Science and Technology of China, Chengdu, 611731, China (e-mail: wmei@uestc.edu.cn; xinwei@std.uestc.edu.cn; boydning@outlook.com).}
}

\maketitle

\begin{abstract}
Movable antennas (MAs) and intelligent reflecting surfaces (IRSs) enable active antenna repositioning and passive phase-shift tuning for channel reconfiguration, respectively. Integrating MAs and IRSs boosts spatial degrees of freedom, significantly enhancing wireless network capacity, coverage, and reliability. In this article, we first present the fundamentals of MA-IRS integration, involving clarifying the key design issues, revealing the performance gain, and identifying the conditions where MA-IRS synergy persists. Then, we examine practical challenges and propose pragmatic design solutions, including optimization schemes, hardware architectures, deployment strategies, and robust designs for hardware impairments and mobility management. In addition, we highlight how MA-IRS integration enhances integrated sensing and communication (ISAC), improving sensing performance and dual-functional flexibility. In general, MA-IRS integration emerges as a compelling approach toward next-generation reconfigurable wireless systems.
\end{abstract}

\section{Introduction}

Future wireless networks demand unprecedented levels of ultra-high capacity, ubiquitous coverage, extreme reliability, and multifunctionality, such as integrated sensing and communication (ISAC). Traditional wireless architectures, constrained by fixed antenna positions and uncontrollable propagation environments, struggle to meet these diverse and stringent requirements. Two emerging wireless technologies, namely movable antennas (MAs) and intelligent reflecting surfaces (IRSs), provide powerful yet complementary solutions, fundamentally redefining how wireless systems can be controlled and optimized. Specifically, MAs actively reshape channel conditions by dynamically repositioning antennas on the transceiver side, thus directly enhancing spatial diversity and beamforming flexibility \cite{zhu2025matutorial}. In contrast, IRSs offer passive yet intelligent manipulation of the wireless environment through reconfigurable reflections, creating favorable propagation paths even under challenging channel conditions \cite{wu2025deployment}. 

Naturally, with effective optimization and accurate channel state information (CSI), integrating these two distinct techniques can yield substantial synergetic gains. Fundamentally, MA-IRS integration considerably boosts spatial degrees of freedom (DoFs), thus improving system capacity, coverage, and resilience against environmental blockage and fading. Such gains are particularly pronounced in complex propagation scenarios, including dense urban or industrial settings. Driven by these theoretical insights, the transformative potential of MA-IRS integration is demonstrated in various wireless scenarios. As illustrated in Fig.~1, representative applications include:
\begin{itemize}
    \item \textbf{Urban cellular networks}, where MA-enabled base stations (BSs) combined with IRS panels mounted in buildings collaboratively provide improved hotspot capacity, multi-area adaptive coverage and seamless urban canyon connectivity without any blind spots.
    \item \textbf{Industrial IoT networks}, where track-based machine-mounted MAs and strategically deployed IRSs jointly ensure robust connectivity in machinery-rich areas, sustain dense sensor networks, and facilitate low-latency robotic coordination.
    \item \textbf{Low-altitude aerial networks}, where integrated aerial and ground-based MA and IRS deployments effectively maintain continuous connectivity and provide flexible dual-function reconfiguration, supporting reliable low-altitude ISAC services such as autonomous transportation and UAV delivery tasks. 
\end{itemize}

However, fully unlocking the potential of MA-IRS integration in real-world wireless networks demands not only an understanding of fundamental design principles but also systematic solutions for practical challenges:
\begin{itemize}
    \item First, effective MA-IRS integration hinges on joint optimization across MA positions, BS beamforming, and IRS phase shifts, all under physical constraints and challenging algorithmic complexity. Realizing these gains requires accurate CSI acquisition, which is hampered by exponential channel state growth from antenna repositioning and extensive pilot overhead for passive IRS elements. Moreover, it remains critical to delineate the scenarios where MA-IRS synergy offers persistent or diminishing returns.
    \item Second, translating the theoretical advantages of MA-IRS integration into practice raises further practical challenges, including CSI overhead, hardware complexity, deployment cost, hardware impairments, and transceiver mobility. To bridge these gaps, pragmatic design solutions are essential, demanding efficient optimization methods with lower CSI estimation overhead, low-complexity and scalable hardware architectures, cost-effective deployment strategies and guidelines, robust designs under hardware imperfections, and mobility management frameworks. 
\end{itemize}
Furthermore, beyond communications, the flexibility of MA-IRS  enables advanced wireless sensing functions, as well as addresses the crucial requirements of ISAC applications. Therefore, the opportunities of MA-IRS integrated systems for ISAC remains to be explored.

Motivated by the above insights and challenges, this article systematically investigates MA-IRS integration from three interconnected perspectives. The main contributions of this work are summarized as follows:
\begin{itemize}
\item We first present the fundamental design issues and principles in MA-IRS integration, quantifying achievable performance gains, including spatial DoFs boost, capacity enhancement, and coverage extension. Crucially, we also delineate specific scenario conditions under which the anticipated MA-IRS synergy either persists or diminishes.
\item Next, we address practical challenges by proposing pragmatic design solutions, including a hierarchical two-timescale optimization framework for CSI overhead mitigation, grouped reconfigurable architectures, e.g., subarray-based MAs and subsurface-based IRSs, for hardware complexity reduction, empirical guidelines for cost-effective deployment strategies, robust system designs accounting for hardware imperfections, and mobility management frameworks for dynamic scenarios.
\item Finally, we explore opportunities for MA-IRS in ISAC applications, emphasizing their distinct and cooperative advantages in sensing coverage, angular accuracy, dual-function flexibility, and mobility adaptability, thereby paving the way for next-generation MA-IRS integrated ISAC wireless systems.
\end{itemize}
Collectively, these analyses establish MA-IRS integration as a compelling paradigm that significantly advances future intelligent, flexible, and reconfigurable wireless networks.

\begin{figure}[!t]
	\centering
	\captionsetup{justification=raggedright,singlelinecheck=false}
	\centerline{\includegraphics[width=0.49\textwidth]{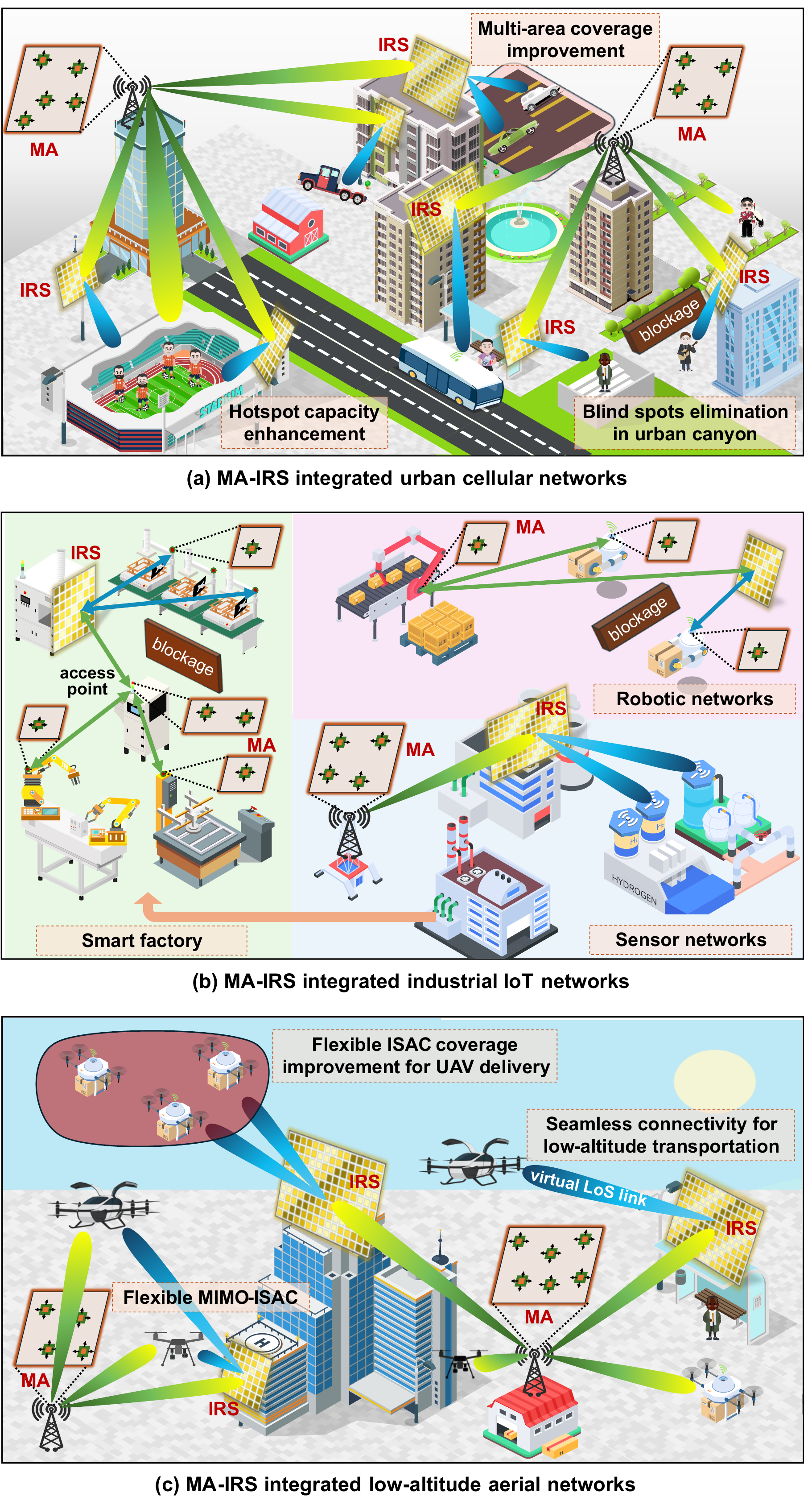}}
	\captionsetup{font=footnotesize}
	\caption{Typical applications for MA-IRS integrated wireless networks.}
	\label{Fig_System}
	\vspace{-15pt}
\end{figure}

\section{Fundamentals of MA-IRS Integration}
The joint use of MAs and IRSs reshapes wireless channels on both the transceiver and the environment sides, unlocking synergetic performance gains. This section first clarifies the design considerations in joint optimization and channel estimation, then quantifies the resulting benefits, and finally identifies the conditions where such synergy persists.

\subsection{Key Design Issues}

\subsubsection{Joint MA position and IRS phase shifts optimization}
Fully exploiting the synergy between MAs and IRSs requires joint optimization across multiple coupled variables: MA positions, BS transmit beamforming, and IRS phase shifts. The related optimization problems are inherently challenging due to the strong coupling between these variables and the resulting high-dimensional, non-convex feasible region. Moreover, practical hardware constraints, such as minimum spacing requirements among MAs to avoid mutual coupling and the unit-modulus constraints enforced on IRS reflection coefficients, further complicate the problem. 
To address these complexities, alternating optimization (AO) has become the predominant framework for tackling joint MA-IRS optimization. The AO approach decomposes the original problem into simpler sub-problems, optimizing each variable set alternately in an iterative manner while keeping the others fixed. Within this AO framework, given fixed MA positions, optimizing transmit beamforming at the BS and reflection patterns at the IRS has been extensively explored. In contrast,  antenna position optimization poses greater practical and theoretical difficulties, typically involving highly non-convex non-concave expressions in both constraints and objective. Most current approaches rely either on gradient-based optimization techniques, such as successive convex approximation (SCA) \cite{zhu2025matutorial}, or evolutionary algorithms, such as particle swarm optimization (PSO) \cite{xiao2024pso}. These methods, though capable of achieving decent optimization performance, suffer from significant computational overhead. An alternative approach discretizes the potential region of antenna mobility into multiple sampling points and optimizes MA positions among these discrete points \cite{wei2025movable}, considerably reducing computational complexity, while achieving performance comparable to gradient-based and evolutionary algorithms.

\subsubsection{Channel estimation}
The effectiveness of joint MA-IRS optimization fundamentally depends on the precise CSI acquisition for all channels involved. However, two practical obstacles severely challenge this requirement. First, traditional IRS-assisted systems, limited by passive reflection elements without signal processing ability, demand extensive pilot overhead to estimate the cascaded BS–IRS and IRS–user channels. Second, the introduction of MAs further complicates the process. Each MA position introduces a new spatial degree of freedom (DoF), causing the number of effective channel states to explode, i.e., with respect to any pair of antenna positions within the transmit/receive regions. Existing channel estimation techniques, designed separately for MA \cite{ma2023compressed, zhang2024channel} or IRS, largely overlook the intertwined characteristics emerging in integrated MA–IRS systems, making them inefficient or even impractical in this joint context.
Therefore, an efficient and generalized channel estimation framework is urgently needed for MA-IRS integrated systems. One promising direction is to exploit the synergy between MAs and IRSs to facilitate CSI acquisition. For example, varying IRS reflection patterns can emulate changes in MA position, modulating received signal strength at the user without physically moving the MA, thereby reducing the need for dense spatial sampling.

\subsection{Performance Gains}

\subsubsection{Spatial DoF boost} A essential benefit of jointly deploying MAs and IRSs lies in their unique ability to boost spatial DoFs, a fundamental determinant of wireless system performance. In this integrated paradigm, MAs dynamically adjust their positions within a specified region, flexibly harvesting spatial diversity and enhancing the system’s ability to combat fading and shadowing effects, while IRSs offer passive beamforming by adaptively controlling the phase shifts of the reflected signals, thus introducing additional tunable propagation paths and manipulating the channel environment. This synergistic enhancement of spatial DoF brings several key usages, such as improving signal strength and link robustness \cite{zhu2025matutorial},\cite{kit2024fasris}, flexible beamforming and interference mitigation, as well as enhanced channel rank and multiplexing capability. In summary, the cooperative exploitation of spatial DoFs by MAs and IRSs transforms the wireless environment from a limited medium into a reconfigurable part of transmission design. The following detail how these DoF enhancements are leveraged for two core system-level gains: capacity enhancement (with an emphasis on MU-MIMO scenarios), and coverage improvement (with a focus on multi-area scenarios).

\subsubsection{Capacity Enhancement} 
In multi-user multiple-input multiple-output (MU-MIMO) systems, the additional spatial DoFs offered by MA–IRS integration translate directly into improved capacity and spectral efficiency. Recent studies demonstrate that enlarging the MA movement region or IRS size yields notable sum-rate gains over conventional systems \cite{sun2025sum,wei2025movable}. Specifically, IRSs generate virtual line-of-sight (LoS) paths, thereby improving channel rank and signal reliability, especially for users at cell edges or in obstructed areas. Meanwhile, MAs further boost performance by adjusting their positions and beam directions to maximize spatial diversity, maintaining high link quality even when direct paths are weak. This synergy is especially beneficial in complex propagation environments, such as urban canyons or dense indoor areas, where IRS reflection paths bypass obstacles, and MA placement collaboratively exploiting these IRS-generated paths, jointly delivering robust connectivity and higher throughput.

\begin{figure}[!t]
	\centering
	\includegraphics[scale=0.65]{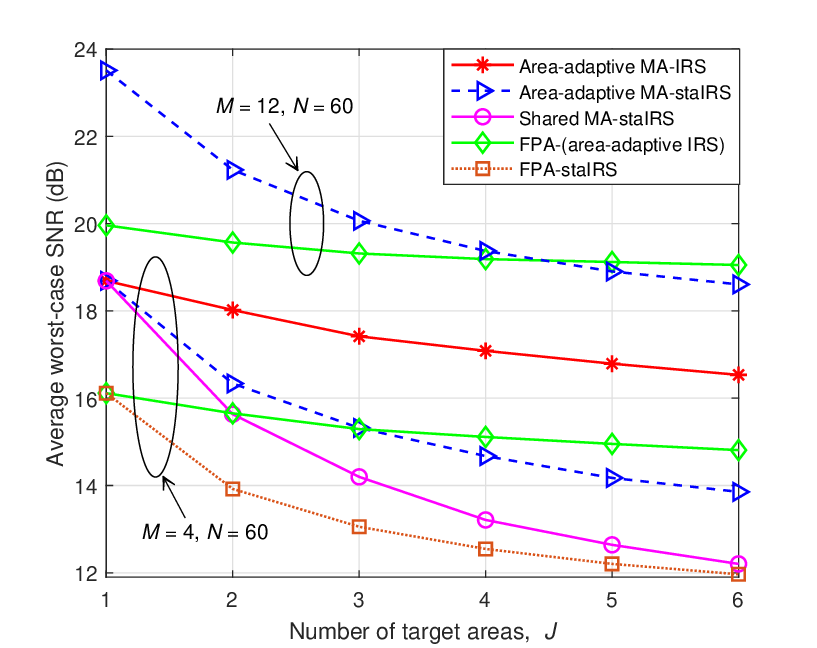}
    \captionsetup{font=footnotesize}
	\caption{Worst-case SNR versus number of target areas under proposed MA-IRS integrated deployments for multi-area coverage improvement.} \label{Fig:SNR_vs_J}
\end{figure}

\begin{figure*}[!t]
	\centering    \captionsetup{justification=raggedright,singlelinecheck=false}
	\centerline{\includegraphics[width=1.0\textwidth]{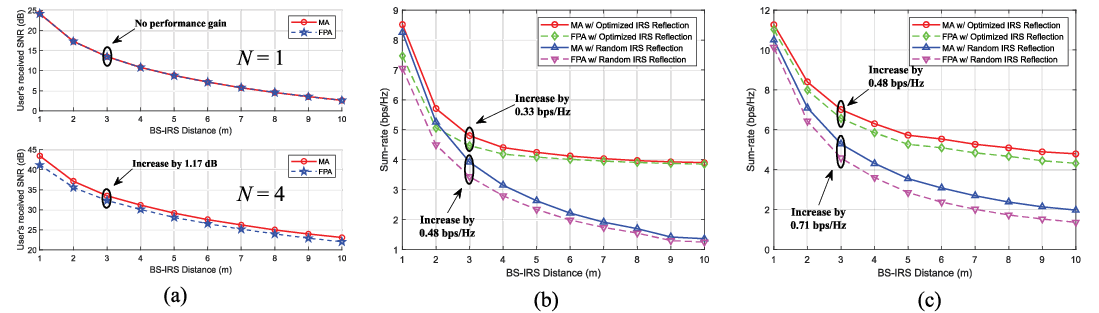}}
	\captionsetup{font=footnotesize}
	\caption{(a) User's received SNR versus the BS-IRS distance under the LoS BS-IRS channel; (b) Sum-rate versus the BS-IRS distance under the LoS BS-IRS channel; (c) Sum-rate versus the BS-IRS distance under the multi-path BS-IRS channel.}
	\label{Fig_Synergy}
	\vspace{-15pt}
\end{figure*}

\subsubsection{Coverage Improvement}
The MA-IRS integration delivers adaptive and scalable coverage, especially valuable for communication blind spots and spatially distributed target areas. There are three representative deployment schemes, with different operational flexibility and cost-effectiveness, to improve multi-area coverage \cite{gao2025coverage}. The first scheme termed the area-adaptive MA-IRS provides the highest DoF, allowing fully dynamic positioning of antennas and IRS reflection configuration tailored to individual coverage regions. A more practical yet still flexible scheme is the area-adaptive MA-staIRS approach, which maintains dynamic MA positioning while deploying static IRSs with fixed reflection configurations determined once during deployment. Finally, the shared MA-staIRS scheme presents the most cost-effective option, employing a common MA placement strategy and static IRS applied across all areas. 

The numerical results depicted in Fig. \ref{Fig:SNR_vs_J} evaluate the worst-case signal-to-noise ratio (SNR) across multiple target coverage areas as their number ($J$) increases. All MA-based schemes consistently outperform their fixed-position counterparts. Notably, the area-adaptive MA-IRS scheme consistently achieves superior performance due to its maximal spatial adaptability. Interestingly, the performance of the area-adaptive MA-staIRS scheme initially surpasses the fixed-position antenna configuration with adaptive IRS, but this advantage progressively diminishes as the number of coverage areas expands, eventually leading to performance crossover. This behavior emerges primarily because the static IRS configurations limit adaptability, failing to adequately enhance channel conditions as coverage complexity grows. In contrast, adaptive IRS configurations, even when coupled with FPAs, retain more robust and stable performance in large-scale coverage scenarios, highlighting the crucial role of adaptability in IRS deployments.
Moreover, increasing the number of MAs from $M = 4$ to $M=12$ amplifies the performance advantage of the area-adaptive MA-stalRS scheme over the FPA-(area-adaptive IRS) scheme in regions where it is superior and narrows the gap where it is inferior, as the additional MAs provide greater spatial diversity to compensate for the limited adaptability of static IRSs.

\subsection{MA-IRS Synergy Conditions}

Despite the promising potential of MA-IRS integration, a critical, yet often overlooked, question concerns the persistence of the performance gains afforded by their synergy. Although extensive research has examined certain joint optimization designs and the resultant performance benefits, due to the comparable capability of MA and IRS for channel reconfiguration, a comprehensive exploration of the mutual interactions between them remains relatively sparse.
To answer this fundamental question, an in-depth investigation into MA-enabled IRS-aided MU-MIMO system, with blocked direct BS-user links and a multi-MA BS, single-FPA user setup, revealed scenario-dependent performance behaviors \cite{wei2025movable}. 

First, for the single-user case, the performance gain of MAs over FPAs may not always exist and depends critically on the BS–IRS channel conditions. Specifically, we assume that the IRS can achieve a LoS-dominant channel with the BS, which usually holds in practice by carefully deploying the IRS in the vicinity of the BS. Then, if the BS-IRS channel follows the near-field propagation, the performance gain of MAs can be observed, as shown in Fig.~\ref{Fig_Synergy}(a). However, in the case of a single-MA BS when $N=1$, the gains vanish even when the MA is located at the nearest position to the IRS. In contrast, for the multi-MA case with $N=4$, when the BS–IRS channel transitions to the far-field regime, the performance gain of MAs gradually diminishes and eventually disappears under the optimal IRS passive beamforming gain. This is because to maximize the received signal power at the user, the BS should set its active beamforming to be aligned with its array response, which eliminates the spatial diversity that MAs could exploit within the movable region. As a result, locating the MA at any position within the movable region yields the identical channel power gain.

Next, for the multi-user scenario, similar limitations apply under specific conditions. In Fig.~\ref{Fig_Synergy}(b), given a LoS BS-IRS channel and typical transmit precoding schemes (e.g., RZF, ZF, MMSE, and MRT), the MAs cannot yield performance gain over FPAs for any given IRS reflection and power allocation \cite{wei2025movable}. This is because typical precoding schemes degenerate into the array response at the BS in the LoS BS-IRS channel, under which the spatial diversity within the movable region diminishes. Furthermore, for the general multipath BS-IRS channel, the performance gap between MAs and FPAs becomes more significant, as shown in Figs.~\ref{Fig_Synergy}(b) and \ref{Fig_Synergy}(c). This indicates that MAs are more beneficial in scenarios with richer scattering, where spatial diversity can be exploited more effectively to enhance communication performance. 
Moreover, Figs.~\ref{Fig_Synergy}(b) and \ref{Fig_Synergy}(c) reveal that MAs yield enhanced performance gains over FPAs when the IRS is randomly configured rather than optimally designed, due to the reduced correlations among the transmit paths in the BS-IRS channel under optimized IRS passive beamforming. In addition, the movable region of the MA is often limited to a wavelength level. Hence, the IRS can leverage its passive beamforming to equalize the channel power gain across all potential MA positions, which reduces the spatial diversity within the movable region, thereby reducing the performance gain of MAs. 
To preserve spatial diversity in MA-enabled systems integrated with IRSs, techniques like mildly sub-optimal or codebook-based IRS patterns, distributed and multi-IRS placements, near-field operation, and six-dimensional MA (6DMA) with full movable degree of freedom may be used, reintroducing spatial variance.

\section{Practical Challenges and Solutions}

Translating the theoretical potentials of MA-IRS integration into practical systems poses several intricate challenges related to CSI acquisition, hardware limitations, and deployment cost. In this section, we examine these practical challenges and present design solutions.

\subsection{Two-Timescale Optimization Scheme}

To manage the overhead and latency issues associated with high-dimensional instantaneous CSI acquisition in practical MA-IRS integrated systems, a hierarchical two-timescale transmission framework emerges as a compelling solution \cite{zheng2024two-timescale}. As shown in Fig. 4, the two-timescale framework decomposes system transmission design into two temporal scales. On the large timescale, the optimization of MA positions and IRS phase shifts leverages long-term statistical CSI, such as spatial covariance matrices, average channel conditions, or user distributions. Decisions at this level focus on achieving objectives such as ergodic throughput maximization or consistent area coverage, which may require updates infrequently, typically on the order of seconds, minutes, or even longer. Such long-term designs reduce computational and estimation overhead by exploiting channel statistics that vary slowly over time. On the small timescale, the BS rapidly adapts its transmit beamforming and power allocation based on instantaneous CSI. With pre-established MA locations and IRS reflection patterns from the larger timescale, the BS maintains real-time adaptability to short-term channel fluctuations, closely matching the agility of conventional beamforming schemes. This two-timescale design reduces the real-time CSI estimation overhead while still capturing the majority of joint MA-IRS optimization benefits. However, while effective in static or quasi-static scenarios, such two-timescale designs face limitations in highly dynamic contexts with rapid channel variations or user mobility. To address this challenge, hybrid strategies combining libraries of pre-optimized MA-IRS configurations and adaptive learning-based algorithms can dynamically select optimal configurations based on real-time environmental changes. 

\begin{figure}[t]
	\centering
	\captionsetup{justification=raggedright,singlelinecheck=false}
	\centerline{\includegraphics[width=0.47\textwidth]{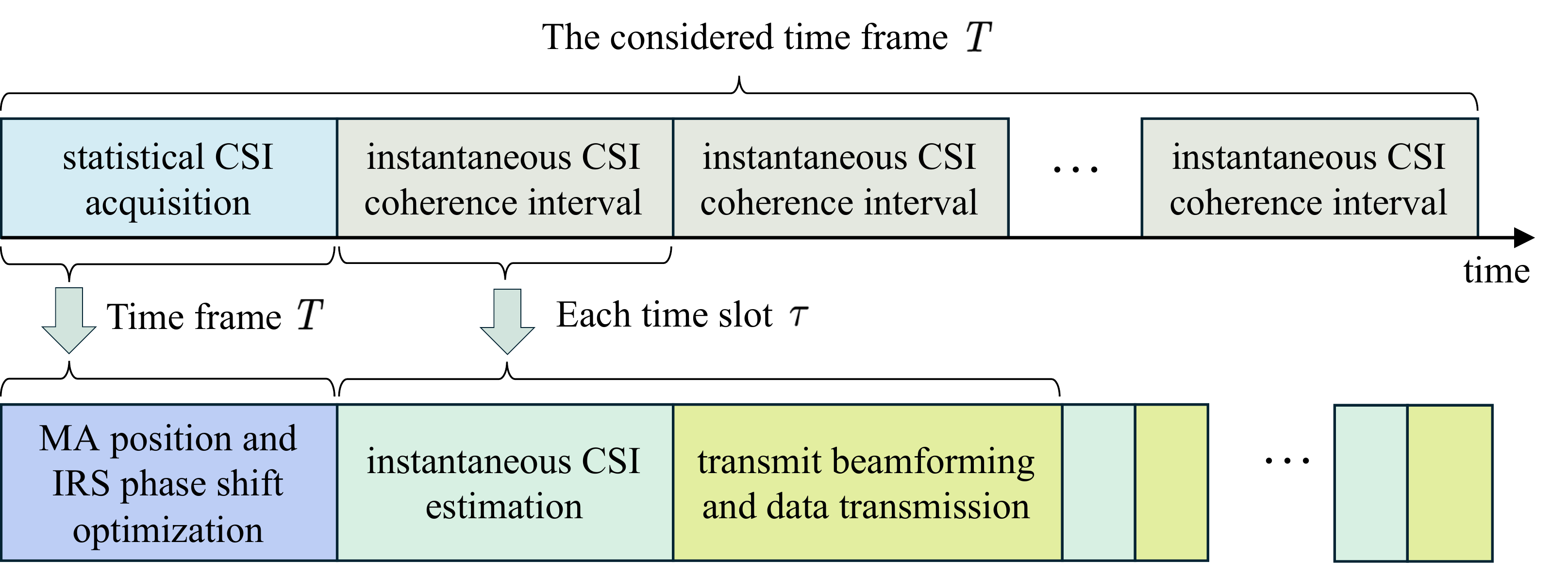}}
	\captionsetup{font=footnotesize}
	\caption{Illustration of a two-timescale MA-IRS optimization framework.}
	\label{Fig_MultiPath_MU}
\end{figure}

\subsection{Subarray and Subsurface Architecture}

In addition to temporal hierarchical designs, another practical consideration is the granularity of reconfiguration in MA-IRS hardware and system control. Although fully element-level phase shift tuning in IRS or antenna-level position adjustment in MAs theoretically maximizes performance, it incurs significant hardware complexity and control overhead, as well as the aforementioned CSI acquisition challenges. By clustering individual IRS elements into subsurfaces or grouping MAs into subarrays, architectures with grouped reconfigurability offer an effective alternative to reduce the number of independently controlled units. For IRSs, the subsurface architecture assigns a common reflection coefficient to each cluster, reducing the CSI acquisition complexity and signaling overhead while preserving substantial beamforming gains, particularly when grouping strategies exploit spatial channel correlations. For MAs, moving antenna panels as collective subarrays substantially reduces mechanical complexity and actuator demands. By optimizing group selection, position, and orientation jointly, such architectures attain near-optimal performance at much lower implementation complexity compared to fully flexible designs. Selecting an appropriate grouping granularity strongly depends on the targeted deployment scenario: highly dynamic or performance-critical scenarios require finer granularity to rapidly adapt and maintain service quality, while larger-scale, cost-sensitive deployments typically benefit from more coarse-grained configurations. Hybrid designs, which combine a few finely movable panels or elements with more coarsely grouped IRSs, often provide optimal trade-offs between performance and complexity.

\subsection{Cost-Effective Deployment Strategy}

System-level deployment in MA-IRS integrated networks involves a trade-off between cost and performance. Although increasing a single type of network components, such as IRS units or MAs, typically enhance overall capability, marginal performance improvement rapidly diminishes beyond certain thresholds due to increased system complexity, hardware expenses, and signaling overhead. Therefore, finding the optimal deployment strategy involves balancing the active spatial diversity gains of MAs against the passive beamforming benefits of IRSs. 

\begin{figure}[!t]
    \centering
    \includegraphics[width=0.48\textwidth]{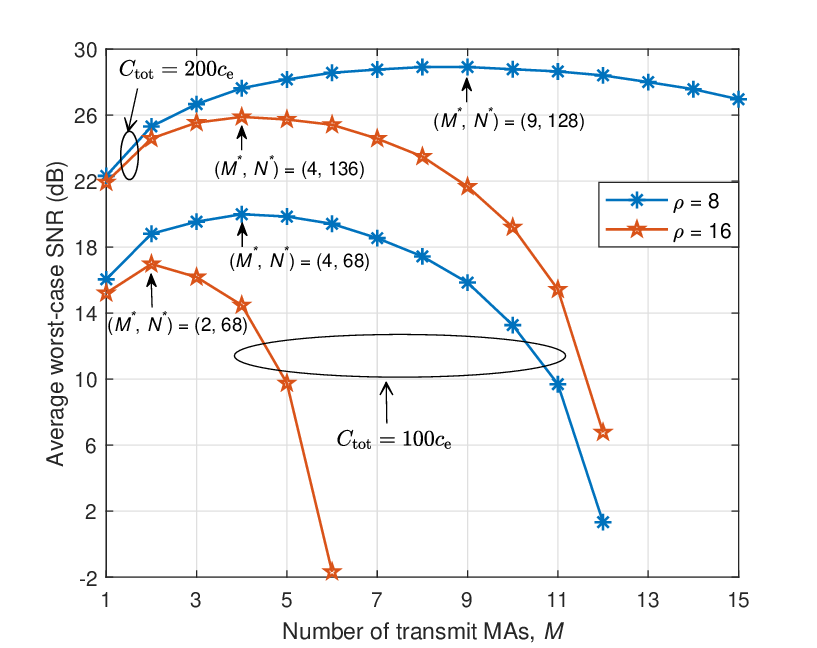}
    \captionsetup{font=footnotesize}
    \caption{Worst‐case SNR of the area‐adaptive MA‐IRS schemes versus number of transmit MAs.}
    \label{dy_vs_M_rhochange}
\end{figure}

To quantitatively explore this balance, we evaluated the worst-case SNR performance of an area-adaptive MA-IRS scheme with a limited total deployment cost in Fig. \ref{dy_vs_M_rhochange}. We define the total cost as $C_{\mathrm{tot}} = \rho c_{\mathrm{e}}M + c_{\mathrm{e}}N$, where $c_{\mathrm{e}}$ is the unit cost per IRS element, $c_{\mathrm{M}} = \rho c_{\mathrm{e}}$ represents the cost per MA, $M$ denotes the number of MAs, $N$ the number of IRS elements, and $\rho$ the cost ratio between a movable antenna and an IRS element. Fig. \ref{dy_vs_M_rhochange} shows the impact of the varying cost ratio $\rho$ and the total budget $C_{\mathrm{tot}}$. An initially increasing SNR is observed as $M$ increases, peaking at an optimal $M^*$, beyond which further increasing $M$ adversely affects performance. It also exhibits a clear non-monotonic trend, reinforcing the existence of an optimal pair ($M^*, N^*$) to achieve the best active-passive resource trade-off. This non-monotonic behavior arises because although increasing $M$ enhances spatial diversity, excessive resource allocation to active MAs at the expense of passive IRS elements undermines passive beamforming capabilities, especially critical in blockage-dominated scenarios relying heavily on IRS reflections. In particular, the optimal MA count $M^*$ is sensitive to two main parameters: it decreases with an increasing cost ratio $\rho$ for fixed budgets, indicating a need to prioritize cheaper IRS elements as MAs become comparatively expensive, and it increases approximately linearly with the total budget for a given cost ratio, signifying that a larger budget accommodates more active MAs without excessively diminishing passive reflection capability.

An empirical relation is identified as: $\frac{M^*}{N^*} \approx \frac{a}{\rho}$, where $a$ encapsulates the balance between active and passive resources. The numerical results indicate $a \approx 0.5$ as a robust baseline for typical scenarios. Under abundant budgets, $a$ slightly increases (e.g., $a\approx0.56$ for $\rho=8, C_{\mathrm{tot}}=200c_{\mathrm{e}}$), whereas in cost-constrained, high-$\rho$ scenarios, $a$ decreases (e.g., $a\approx0.47$ for $\rho=16, C_{\mathrm{tot}}=100c_{\mathrm{e}}$). Crucially, this empirical relation demonstrates that the optimal ratio between active MAs and passive IRS elements predominantly depends on the relative cost ratio $\rho$, rather than the absolute total budget. Practically, combining this empirical insight with the total cost constraint yields a simplified design guideline: $M^*\approx\frac{1}{3}\frac{C_{\mathrm{tot}}}{\rho c_{\mathrm{e}}}$, providing a convenient yet accurate approach to selecting the optimal number of movable antennas for given budget constraints.

\subsection{Robust Design under Hardware Imperfections}

The practical implementation of MA-IRS systems with favorable performance gains may be hampered by various hardware imperfections. For IRSs, limitations such as finite phase resolution, amplitude-phase coupling, switching delays, phase noise, and mutual coupling all contribute to deviations from idealized reflection behaviors. For MAs, actuator precision limits, position tuning inaccuracies, mechanical delays \cite{wang2024delay}, and mobility constraints significantly impair the system's ability to optimally exploit spatial diversity. Crucially, these imperfections rarely occur in isolation and often mutually amplify performance degradation. For example, MA positioning errors alter the channel response between MA and IRS, rendering optimized reflection patterns suboptimal, while IRS phase inaccuracies mislead optimal antenna movement decisions. Such joint effects create a non-linear propagation of errors and compounded performance losses throughout the system, severely complicating robust system design and also accurate CSI estimation. In practice, two main approaches can mitigate these challenges. First, comprehensive calibration schemes, such as embedded sensors for self-calibration of IRS phase shifters and real-time monitoring of MA positions, facilitate joint estimation and compensation of these coupled impairments from the hardware level. Second, from the algorithm level, robust algorithmic designs that explicitly account for hardware imperfections, including statistical or bounded error models, and the development of codebooks and beam patterns resilient to phase and positioning errors, can help mitigate system performance degradation in practice.

\subsection{Mobility Management Framework}

The dynamic repositioning capabilities of MAs combined with agile IRS reflection adjustments can be potentially advantageous in high-mobility scenarios, such as low-altitude UAV communications and autonomous vehicle communications. However, rapid user or platform movement leads to fast channel variations and shortened coherence intervals, making real-time MA and IRS adjustments for mobility management a key practical challenge. 
Addressing these challenges requires advanced hierarchical mobility management frameworks. After establishing an initial channel estimate using structured pilots with element grouping strategies, predictive tracking algorithms, such as Kalman filtering, particle filtering, or deep learning-based sequence models, continuously update the channel state by leveraging spatio-temporal correlations. Real-time planners employing reinforcement learning or graph optimization strategies then translate these predictions into optimal MA trajectories and IRS configurations, dynamically adjusting antenna positions and reflection patterns. Such predictive mobility-aware strategies ensure sustained link quality and connectivity even under rapid environmental changes, turning mobility from a fundamental bottleneck into a powerful operational asset. In addition, for UAV- or vehicle-mounted arrays, MA trajectory optimization must also jointly consider flight dynamics, coverage requirements, and link budgets to dynamically select optimal antenna positions; IRS beam tracking can benefit from a two-stage approach that combines slow codebook-based adaptation with fast local refinements.

\section{MA-IRS Integration for ISAC}

The synergies and challenges of MA-IRS integration in the above discussions also align closely with the core performance dimensions of wireless sensing, as well as ISAC applications, namely, sensing coverage, angular resolution, beam agility, and latency. This section revisits the MA-IRS paradigm beyond communications through a sensing and ISAC perspective.

\subsection{Advantages in Wireless Sensing}
MA can adjust its position to exploit spatial diversity, actively performing sensing by changing the array geometry. In contrast, IRS passively steers reflection beams to redirect signal energy, reshaping the sensing environment. When integrated, these two complementary functionalities deliver superior sensing coverage and enhanced accuracy, particularly in suppressing coverage holes and boosting sensing performance.

\subsubsection{Sensing Coverage Enhancement}
A prominent obstacle in wireless sensing, particularly in cluttered or indoor environments, is the difficulty of reliably detecting and localizing targets obstructed by blockage, commonly known as non-LoS (NLoS) conditions. MA-IRS integrated systems tackle this challenge cooperatively. Specifically, IRS is utilized to dynamically generate virtual LoS paths by steering reflected signals toward obstructed targets. Note that the effectiveness of this IRS-enabled strategy critically depends on optimal IRS placement and reflection pattern configurations. Complementing IRS capabilities, MAs actively enhance sensing performance by repositioning antennas to amplify signal strengths received via IRS-generated virtual paths. This collaborative approach creates multiple diverse propagation routes, robustly penetrating obstructed or shadowed areas. In this way, MA-IRS integration can significantly extend sensing coverage and mitigate detection blind spots, improving the reliability and robustness of sensing tasks in complex environments ranging from urban canyons to dense indoor facilities.

\subsubsection{Sensing Accuracy Improvement}

Angular resolution, determined largely by the physical aperture size of the antenna arrays, is crucial for accurate target detection and localization. However, deploying large, dense antenna arrays to improve angular resolution can be impractical due to cost, space, and energy limitations. The MA-IRS integrated architecture addresses this constraint by synthesizing large effective apertures through coordinated MA repositioning (reshaping geometry as a sparse array) and distributed IRS deployment. This integrated approach dramatically improves angular resolution without necessitating prohibitively dense active antenna deployments.
Furthermore, joint optimization of MA positions and IRS phase shifts can significantly lower the achievable Cramér-Rao bound (CRB), a theoretical lower bound on estimation accuracy, thereby enhancing sensing precision. Mechanically, CRB reduction in MA–IRS sensing obeys a cooperative behavior of power and geometry: IRS-driven coherent reflections increase the received SNR, whereas MA position/orientation enlarges the effective aperture and sensitivity of the estimation parameter in the Fisher information matrix. Consequently, combined reconfigurability, offered by optimizing MAs to maximize spatial variance and employing IRS elements to deliver high SNR reflection paths from diverse directions, substantially improves both angular and positional estimation accuracy \cite{ma2024movable}. 

\subsection{Advantages in ISAC}

ISAC leverages shared wireless resources to simultaneously support data transmission and environmental sensing, enhancing spectral efficiency, and enabling advanced functionalities such as joint radar-communication systems. A central challenge arises from balancing the distinct beamforming needs of these functions: communication typically demands user-specific directional beams, while sensing benefits from broader coverage or finer spatial resolution. The integration of MA and IRS significantly expands spatial and configurational flexibility. By jointly optimizing antenna positions and IRS reflection patterns, MA-IRS architectures effectively manage communication-sensing trade-offs, resulting in adaptive ISAC solutions. In the following, we discuss two key capabilities enabled by MA-IRS integration: dual-function beamforming and robust mobility management.

\begin{figure}[t]
    \centering
    \includegraphics[width=3.3in]{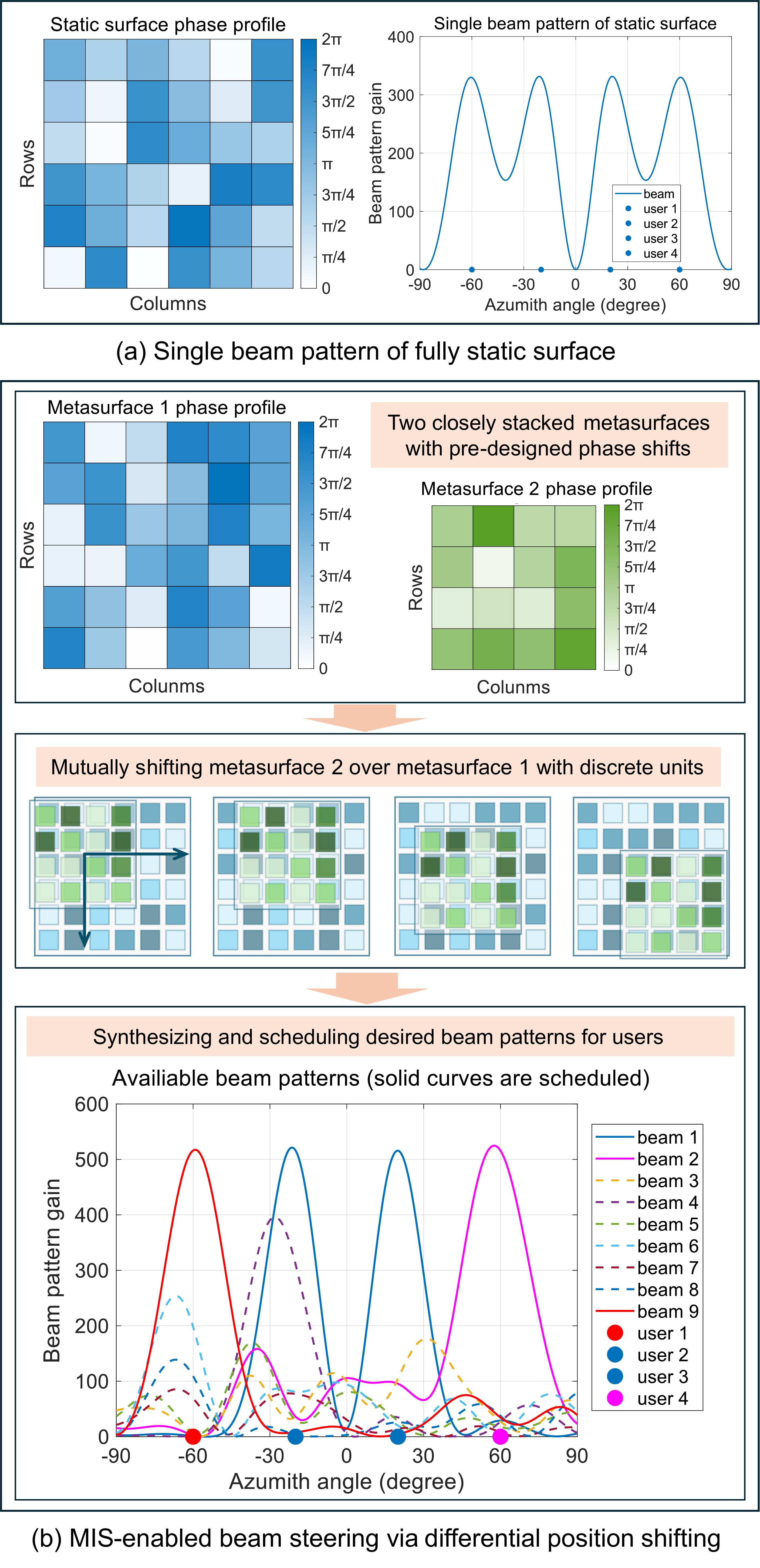}
    \captionsetup{font=footnotesize}
    \caption{An example of the MIS beam pattern switching mechanism under a max-min user SNR rule: (a) the phase profile of a single-layer static surface and its resulting single beam pattern; (b) by sliding a small pre-phased MS2 over a larger static MS1, the MIS creates multiple composite phase profiles, and hence multiple beam patterns, without electronic tuning. All possible beams are plotted, and the solid curves show those selected for four users with significantly improved beam gain compared to case (a).} 
    \label{fig:system_model}
    \vspace{-12pt}
\end{figure}

\subsubsection{Flexible Dual-Function Reconfiguration}

A unified ISAC framework demands beamforming strategies that can harmonize different requirements. The MA-IRS synergy notably expands spatial controllability via joint optimization of antenna positions, transmit beamforming, and IRS phase shifts, simultaneously maximizing communication throughput and sensing accuracy \cite{wu2025movable}. Further potential stems from dynamic spatial-temporal reconfigurability in MA-IRS systems. Spatially or temporally partitioned IRS configurations, combined with real-time antenna repositioning, allow the system to flexibly adjust beam patterns, using narrow beams for focused communication while simultaneously maintaining wide-angle beams for sensing targets. Practical implementations include temporarily repositioning antennas to emulate a larger virtual aperture for improved sensing resolution, subsequently reverting to communication-optimized layouts. Such dynamic reconfigurability enables flexible communication capacity and sensing precision enhancement on demand.

\subsubsection{Robust Mobility Adaptation}

High-mobility scenarios such as vehicular networks (V2X), drone-based sensing, and high-speed trains pose stringent demands due to rapid channel variations, Doppler effects, and frequent signal obstructions. MA-IRS integration offers critical adaptability to cope with these challenges. Specifically, MAs provide spatial flexibility by repositioning antennas to maintain optimal LoS or favorable sensing conditions, while IRS elements quickly redirect reflected beams to sustain communication links and sensing coverage. This dual reconfigurability allows MA-IRS-equipped systems to track mobile targets, quickly compensate for obstructions, and effectively mitigate rapid channel fluctuations. Moreover, the MA-IRS integrated approach enables predictive mobility management and proactive Doppler mitigation by leveraging predictive algorithms such as Kalman filtering or deep learning-based trajectory estimation. MA-IRS systems can dynamically adjust antenna positions and IRS configurations to minimize relative velocity and angle variations, reducing Doppler-induced performance degradation. Furthermore, IRSs can maintain stable reflective paths during intermittent LoS interruptions, complemented by MA geometric adjustments within physically constrained motion paths.

\section{Conclusion and Future Directions}

This article examined the integration of MAs and IRS, highlighting their synergy in boosting spatial DoFs to improve capacity, coverage, and sensing performance in wireless networks.
We first clarified the fundamental principles, key design issues, and complementary capabilities of MAs and IRSs, quantitatively revealing the performance gains and synergy conditions enabled by their joint deployment. 
Then, to bridge theory and practice, we provided pragmatic system design solutions addressing key practical challenges, including the two-timescale optimization, subarray and subsurface architecture, cost-effective deployment, robust design, and mobility management.
Furthermore, MA-IRS integrated systems show unique advantages for ISAC, delivering improved sensing coverage, angular resolution, and flexible dual-function reconfiguration. Collectively, these findings establish MA-IRS integration as a promising enabler for next-generation reconfigurable wireless networks.

In the future, several promising research directions can further unleash the potential of MA-IRS integration.  First, the development of advanced movable IRS architectures, rather than separately implemented MAs or IRSs, such as flexible intelligent metasurfaces (FIM), subsurface-wise movable IRS, panel rotation-based six-dimensional MA (6DMA) \cite{shao20256dma} can enable enhanced fine-grained beam control for enhanced capacity and flexible coverage; meanwhile, as movable intelligent surfaces (MIS) \cite{zheng2024MIS}, operated with differentially position shifting mechanism depicted in Fig. 6, effectively eliminates the reliance on electronic tuning incurred by massive phase elements, thereby significantly reducing the hardware complexity and energy consumption, while still providing beam steering capabilities for quasi-static wireless networks. Second, the application of artificial intelligence and machine learning for real-time adaptive control of MA and IRS parameters represents a promising avenue. Such approaches are expected to dramatically reduce the CSI acquisition overhead and facilitate efficient and robust operation in highly dynamic or uncertain wireless environments. Third, advancing high-frequency MA-IRS integration is essential. Future research should focus on channel modeling, beamforming, and deployment strategies specifically tailored for the mmWave and THz bands. This includes addressing near-field and hybrid near-far-field effects, stringent alignment requirements, and wideband challenges such as beam-squint.

\ifCLASSOPTIONcaptionsoff
  \newpage
\fi



%

\section*{Acknowledgment}
Q. Wu's work is supported by the National Key R\&D Program of China (2023YFB2905000), the National Natural Science Foundation of China (NSFC) 62371289, NSFC 62331022, and Shanghai Jiao Tong University 2030 Initiative. W. Mei's work is supported by the Natural Science Foundation of Sichuan Province under Grant 2025ZNSFSC0514. W. Chen's work is supported by NSFC 62531015 and Shanghai Kewei 24DP1500500. Y.~Gao's work is supported by the NSFC under Grant 62501389.

\section*{Biographies}

Qingqing Wu is an Associate Professor at Shanghai Jiao Tong University. His current research interests include intelligent reflecting surface (IRS), uncrewed aerial vehicle (UAV) communications, and MIMO transceiver design. He has co-authored over 100 IEEE journal papers, including more than 40 ESI highly cited papers, which have received over 45,000 Google Scholar citations. He has been listed as a Clarivate ESI Highly Cited Researcher since 2021.

Ziyuan Zheng received B.Eng. and Ph.D. degrees from the School of Information and Communication Engineering, Beijing University of Posts and Telecommunications (BUPT), Beijing, China, in 2018 and 2024, respectively. He is currently with Shanghai Jiao Tong University (SJTU) as a post-doctoral researcher. His main research interests include reconfigurable intelligent surface (RIS), movable antenna (MA), and satellite communications (SatCom).

Ying Gao received the B.Eng. degree from Nanjing University of Science and Technology (NJUST), China, in 2016, and the Ph.D. degree from the Shanghai Institute of Microsystem and Information Technology (SIMIT), Chinese Academy of Sciences (CAS), in 2021. From 2021 to 2023, she was a Post-Doctoral Researcher with the University of Macau (UM). Since December 2023, she has been a Post-Doctoral Researcher with the Department of Electronic Engineering, Shanghai Jiao Tong University (SJTU). Her research interests include intelligent reflecting surfaces, movable antennas, unmanned aerial vehicle communications, physical-layer security, and optimization theory.

Weidong Mei (wmei@uestc.edu.cn) is currently a Professor with the National Key Laboratory of Wireless Communications, University of Electronic Science and Technology of China. His research interests include reconfigurable antennas, intelligent reflecting surfaces, wireless drone communications, and convex optimization techniques. Dr. Mei has been listed in World's Top 2\% Scientists by Stanford University since 2021. He was the recipient of the Best Paper Award from the IEEE International Conference on Communications in 2021. He was honored as the Exemplary Editor of the IEEE Open Journal of the Communications Society in 2024.

Xin Wei (Graduate Student Member, IEEE) received the B.S. degree in communication engineering from Chongqing University of Posts and Telecommunications in 2023. He is currently pursuing the M.S. degree in information and communication engineering with the National Key Laboratory of Wireless Communications, University of Electronic Science and Technology of China. His research interests include movable antennas, wireless drone communications, and convex optimization. He was the co-recipient of the IEEE ComSoc SPCC Technical Committee Student Challenge and Video Contest Award in 2024.

Wen Chen [M’03, SM’11] is now a tenured Professor at the Department of Electronic Engineering, Shanghai Jiao Tong University, China. He is the Shanghai chapter chair of the IEEE Vehicular Technology Society. His research interests include green multiple access, wireless AI, and RIS communications. He has published more than 200 papers in IEEE journals and has over 11,000 citations in Google Scholar. 

Boyu Ning (Member, IEEE) received the B.S. degree in communication engineering from the Yingcai Honors College, University of Electronic Science and Technology of China (UESTC), Chengdu, China, in 2018, and the Ph.D. degree from the National Key Laboratory of Wireless Communications, UESTC, Chengdu, China, in 2023. From 2022 to 2023, he was a Visiting Student with the Department of Electrical and Computer Engineering, National University of Singapore. His research interests include terahertz communication, movable antennas, intelligent reflecting surface, massive MIMO, physical-layer security, and convex optimization.

\end{document}